\documentclass[twocolumn,aps,prb,graphicx,showpacs]{revtex4}
\usepackage{amssymb}
\usepackage{epsfig}
\usepackage{graphicx}
\usepackage{amsfonts}
\usepackage{amsmath}
\usepackage{epstopdf}

\begin{document}

\title{Semiclassical analysis  of the Nonequilibrium Local Polaron}
\author{A. Mitra, I. Aleiner and A. J. Millis}
\affiliation{Department of Physics, Columbia University, 538 W. 120th St, NY, NY 10027}
\date{\today}

\begin{abstract}
A resonant level strongly coupled to a local phonon under nonequilibrium
conditions is investigated. The
nonequilibrium Hartree-Fock approximation is shown to correspond to
approximating the steady state density matrix by delta functions at field
values to which the local dynamics relaxes in a semiclassical limit. If
multiple solutions exist, all are shown to make nonvanishing contributions to
physical quantities: multistability does not exist. Nonequilibrium effects
are shown to produce decoherence, causing the standard expansions to
converge and preventing the formation of a polaron feature in the spectral
function. The formalism also applies to the nonequilibrium Kondo problem.
\end{abstract}

\pacs{73.23.-b,05.30.-d,71.10-w,71.38.-k}
\maketitle

The quantum mechanics of nonequilibrium systems is a subject of fundamental
importance and of great current interest, for example in the context
of prospective 'single molecule devices' \cite{Molecule}. In equilibrium problems, nonperturbative
analysis based on solutions of Hartree-Fock equations
(which may be understood as saddle points of functional integrals)
has led to important insights.  In nonequilibrium problems, mean field equations
may be formulated \cite{Kamenev,Alexandrov02,Gogolin04} and indeed exhibit a
richer structure than in the corresponding equilibrium problems, but it has
not been clear how to select the relevant solutions or to systematically
compute  corrections.

In this paper we investigate a simple model which indicates
a resolution to  these issues. We find that the relevant solutions
are selected by the steady state density matrix, which in
a  semiclassical, weakly nonequilibrium limit
is found to become very sharply peaked at field values corresponding
to local minima of  a 'pseudoenergy' which we define. The formalism also shows that
if multiple minima exist, all contribute, with weights varying smoothly
as parameters change and shows 
how departures from equilibrium lead to a decoherence which 
suppresses characteristically quantal effects such as the formation of a polaron resonance.

We consider a single level which may be occupied by $0$ or 
$1$ spinless electrons (creation operator $d^\dagger$) and
is coupled to two leads $j=L,R$ (creation operators $a^{\dagger}_{jk}$).
The leads are assumed to be
reservoirs specified by chemical potentials $\mu _{j}$ and inverse
temperature $\beta _{j}$ (which we typically set to $\beta _{j}=\infty $).
The electrons interact with an oscillator (dimensionless displacement coordinate $q$
and momentum $p$) of
mass $M_{ph}$ and energy $U(q)$). This is a local version of the
familiar 'polaron problem' \cite{Mahan4} and captures an important aspect of
the physics of prospective single-molecule devices \cite{McEuen00,Mitra04}. 
The Hamiltonian is $H=H_{ce}+H_{tun}+H_{dot}+H_{ph}+H_{I}$; with 
\begin{equation}
H_{ce} =\sum_{k,j}\epsilon _{k,a}a_{k,j}^{\dagger }a_{k,j}\text{; }
H_{dot}=\epsilon _{0}d^{\dagger }d\text{;}  \label{hce} 
\end{equation}
\begin{equation}
H_{tun} =\sum_{k,j}t_{k,j}\left( d^{\dagger }a_{k,j}+h.c\right) 
\label{htun}
\end{equation}
\begin{equation}
\text{ }H_{I} =\lambda qd^{\dagger }d\text{; \ }H_{ph}=\frac{p^{2}}{2M_{ph}}+U(q)
\label{hph}
\end{equation}

\begin{figure}[th]
\includegraphics[width=3.0in]{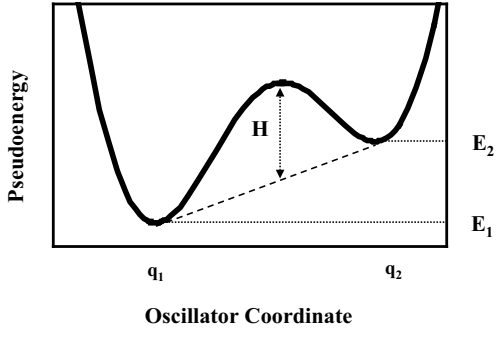}
\caption{Assumed dependence of 'pseudoenergy' (Eq. \ref{Phi}) on
oscillator coordinate $q$.}
\label{fig:E}
\end{figure}

At $M_{ph}=\infty$ 
the model is analytically solvable, with properties
determined by the Green functions
\begin{equation}
\mathbf{g_q}(t-t^{\prime})=%
\begin{pmatrix}
{g}_{qR}(t-t^{\prime}) & {g}_{qK}(t-t^{\prime}) \\ 
0 & {g}_{qA}(t-t^{\prime})%
\end{pmatrix}
\label{gdef}
\end{equation}
where $g(t)=\int\frac{d\omega}{2\pi}e^{-i\omega t}g(\omega)$ with 
\begin{eqnarray}
{g_q}_{R}(\omega )& =\frac{1}{\omega -\epsilon _{0}-\lambda q-\Sigma (\omega
+i\delta )}={g}_{qA}(\omega )^{\ast }  
\label{grdef} \\
{g}_{qK}(\omega )& =2i\sum_{j=L,R}
\left[ a_{j}(\omega )\tanh (\beta_j(\omega -\mu _{j}))\right] ,  
\label{gkdef}
\end{eqnarray}
and $\Sigma =\Sigma _{L}+\Sigma _{R}$, $\Sigma _{j}(z)=\sum_{p}\frac{t_{p,j}^{2}%
}{z-\epsilon _{pa}}$ and $a_{j}=Im\Sigma _{j}(\omega -i\delta )/\left(
\left( \omega -\epsilon _{0}-\lambda q-Re\Sigma (\omega )\right) ^{2}+
Im\Sigma ^{2}(\omega )\right) $. It is also useful to define $b=Reg_{R}$
and the 'pseudoenergy'
\begin{equation}
\Phi(q)=U(q)+\int_0^q dq'\lambda q'\int \frac{d\omega}{2\pi i}\frac{g_{q'K}-g_{q'R}+ g_{q'A}}{2}
\label{Phi}
\end{equation}
which in equilibrium becomes the q-dependent part of the total energy.
Fig. \ref{fig:E} shows a possible form of $\Phi (q)$. 

The equilibrium physics of this model is very well understood, and is
conveniently viewed as a functional integral
over trajectories $q(t)$.    In the limits $M_{ph}$,$\beta \rightarrow \infty $
the integral is dominated by those paths for which $q$ takes the definite
value $q_{a}$ minimizing the energy. At finite $M_{ph}$, other
paths become important: two crucial classes are the 'gaussian fluctuation'
paths involving small excursions (characteristic frequency $\omega_a$) from
the miniman and,  if $\Phi _{eq}(q)$
has the form shown in Fig \ref{fig:E} and the barrier height
$H$ is sufficiently large,  tunnelling (instanton) processes  during which $q$ goes rapidly  from
the vicinity of the global minimum ($E_{1}$ in the notations of Fig \ref{fig:E}) to
the higher energy minimum (here $E_{2})$, spends a time of order $\Delta
E^{-1}\equiv \left( E_{2}-E_{1}\right) ^{-1}$ near the higher minimum, and
then returns to the vicinity of the global minimum. If $H$ is sufficiently
large, then tunnelling processes are rare, but when $E_{2}\approx E_{1}$,
the time spent in the higher minimum becomes longer than the interval
between tunnelling events and the dilute instanton approximation breaks
down, signalling the formation of a polaron resonance in the density of
states. In this paper we investigate the changes occurring when the system is driven
out of equilibrium by application of a chemical potential difference $\Delta
\mu $ between the two leads. We present a general discussion but focus most
attention on the nonequilibrium polaron limit $H>>\omega _{a}>>\Delta \mu $;
with $\beta_j =\infty $ but $\Delta E/\Delta \mu $ arbitrary. In our actual
calculations we also assume that departures from equilibrium are small enough
that we may neglect density-of-states variations: $(\Delta \mu) \partial ln a /\partial\omega<1$.

To analyse the out of equilibrium behavior we use the standard \cite{Keldysh63,Rammer}
extension of Feynman diagrammatics to the \textit{%
Keldysh contour} which consists of a time-ordered ($-)$ branch extending
from $t=-\infty $ to $t=+\infty $, followed by an anti-time-ordered ($+$)
branch extending from $+\infty $ back to $-\infty $. The Keldysh
diagrammatics may be obtained by functional differentiation, with respect to
source terms $\eta _{a}(t)$, of a generating functional $W[\{\eta _{a}]$
which can be formulated \cite{Kamenev} as a coherent
state path integral:
\begin{equation}
W[\{\eta _{a}\}]=\int dq _{+}dq _{-}\hat{\rho}(q _{+},q
_{-})\int \mathcal{D}q (t)\hat{S}_{K}[-\infty _{+},-\infty _{-}]
\label{Wpath}
\end{equation}%
Here the functional integral $\int \mathcal{D}q (t)$ is over all
paths beginning at $q _{-}$ at $t=-\infty $ on the time ordered contour
and ending at $q _{+}$ at $t=-\infty $ on the antitime ordered contour, while
$\hat{S}_{K}[t_{1},t_{2};\eta_{a}(t)]=T_{K}e^{i\int_{t_{2}}^{t_{1}}%
dta(t)\hat{H}[\eta _{a}(t)]}$ the time evolution operator on the Keldysh contour
and $a(t)=\pm 1$ according to
whether the time is on $-$ or $+$ branch.  
The contributions of paths with endpoints $q _{+},q _{-}$ are weighted
by the appropriate element of the steady state
density matrix $\rho $ which is the long time
limit of an equation which 
may be expressed in path integral language as
\begin{eqnarray}
\widehat{\rho }(q _{+},q _{-};t;\eta )& =\int dq _{+}^{\prime
}dq _{-}^{\prime }\int \mathcal{D}^{\prime }q ^{\prime }(t^{\prime })
\label{rhooft} \\
& \hat{S}_{K}[t_{+},t_{0+}]\widehat{\rho }(q _{+}^{\prime },q
_{-}^{\prime };t_{0})\hat{S}_{K}[t_{0-},t_{-}]  \notag
\end{eqnarray}%
Here the $\int \mathcal{D}^{\prime }q
^{\prime }(t^{\prime })$ is over all paths which begin at $q _{-}^{\prime
}$ on the lower contour at time $t_{0}$ and end at $q _{-}$ time $t$ on
the lower contour, along with the time reversed paths which begin at $q
_{+}^{\prime }(t)$ on the upper contour and return to $q _{+}$ at time $%
t_{0}$.  We require the long time behavior after transients have decayed. We
do not find non-steady long time limits such as limit cycles or chaos:
$\rho $ in Eq \ref{Wpath} is the time independent solution of Eq \ref%
{rhooft}.

Because eqs \ref{hce},\ref{htun},\ref{hph} involve a finite system coupled
to two infinite reservoirs, the trace over electron operators may be
performed \cite{Mitra04}.  The combination of time evolution operators
needed in eqs. \ref{Wpath},\ref{rhooft} 
may be written in the interaction representation as
${\cal R}(t_2,t_1;\left\{q(t)\right\})\equiv\langle T_K Exp[-i\int_{t_1}^{t_2}dt(\lambda q_c(n_--n_+)+\lambda_qq_q(n_-+n_+))]\rangle$
where $q_{c}(t)=\frac{q_{-}(t)+q_{+}(t)}{2}$, $q_{q}(t)=\frac{q_{-}(t)-q_{+}(t)}{2}$, 
$\langle \rangle$ denotes expectation value in the
reservoir defined by the leads (note this trace depends on the entire
trajectory $q(t)$) and $n_{+/-}=d^{\dagger}d$ on the indicated contour.
In the physical problem
the quantum field coupling constant $\lambda_q=\lambda$.  Differentiation with 
respect to $\lambda_q$
and application of the usual linked cluster arguments leads to 
\begin{equation}
{\cal R}(t_2,t_1;\left\{q_c(t),q_q(t)\right\})=e^{-\int_{t_{1}}^{t_{2}}dt(C(t)+\mathcal{L}_{ph+S})} 
\label{Slinked}
\end{equation}%
Here $\mathcal{L}_{ph+S}$ is the phonon action 
$M_{ph}(\partial_t q_q \partial_t q_c)-U(\frac{q_q+q_c}{2})-U(\frac{q_q- q_c}{2})$
supplemented by source terms, 
\begin{equation}
C(t)=q_{q}(t)\int_{0}^{\lambda}d\lambda_q Tr\left[\mathbf{\sigma_x G}_{\lambda_q}(t,t)\right]
\label{Cdef}
\end{equation}%
and $\mathbf{G}_{g^{\prime }}$  solves
\begin{equation}
\mathbf{G}_{\lambda_q}=\mathbf{g}+\mathbf{g}
\mathbf{\ast v\ast G}_{\lambda_q}  
\label{glamdef}
\end{equation}
with  $\mathbf{v}(t)=\left( \lambda q_{c}(t)\mathbf{1}+\lambda_qq_{q}(t)%
\mathbf{\sigma }_{x}\right) $. We have verified that  an
expansion in powers of $\lambda$ about the weak coupling limit $\lambda=0$
reproduces results obtained \cite{Mitra04} by standard Keldysh diagrammatic analysis. 

We now turn to the
semiclassical analysis.  In the large $M_{ph}$ limit, one expects stationary
(time-independent) paths to dominate the physics.  However, in the Keldysh
formalism, stationary paths have $q_{q}=0$,
so in the absence of source terms the time evolution operator in Eq \ref%
{Wpath} is unity for all stationary paths, providing no basis for
selection.  Instead, the important paths are selected by the
density matrix; i.e. from the solution of Eq \ref{rhooft}.  
%A key role
%in this solution is played by the 'pseudoenergy' $\Phi$ defined in Eq \ref{Phi}.
%Many different cases can arise; here we take
%$\Phi (q)$ to have the form
%shown in Fig ~\ref{fig:E}. 

To understand the dynamics implied by Eq \ref{rhooft}
we first assume that $\rho (q,q^{\prime })$ is strongly peaked near $q=q^{\prime }=q_{a}$. 
We may then
expand $\Phi =E_{a}+\frac{M_{ph}\omega _{a}^{2}}{2}\left(
q-q_{a}\right) ^{2}$ and analyse Eq \ref{rhooft} by the usual perturbative
methods \cite{Mitra04}.  
In the large mass, weakly nonequilibrium limit we find
relaxation, with a rate of order $\lambda^{2}\omega _{a}(a_{L}+a_{R})^{2}$ to a sum of
functions sharply peaked about the values $q_a$ which minimize the pseudoenergy.
\begin{equation}
\rho(q,q') \approx \sum_a \rho_a r_a(q,q')
\label{rhoapprox}
\end{equation}
To leading order in the electron-phonon coupling $\lambda$
we find $r_a(q,q')=\int\frac{dp}{2\pi}e^{ip(q-q')}
Exp[-(\frac{p^2}{2M_{ph}}+\frac{M_{ph}\omega_a^2(\frac{q+q'}{2}-q_a)^2}{2})
\frac{2tanh[\frac{\omega_a}{2T_{eff}}]}{\omega_a}]$ with 
$T_{eff}=\Delta \mu \frac{a_{L}a_{R}}{\left( a_{L}+a_{R}\right)
^{2}}$, peaked about $q_a$ with a width arising from
quantum fluctuations (finite $M_{ph}$) and from departures
from equilibrium (which act as an effective temperature).
%The full expressions are cumbersone, but in
%the limit $\Delta \mu <\omega_a$ $r_a(q,q^{\prime })=
%\frac{1}{\sqrt{\pi \xi _{a}}}
%e^{-\left( \left(q-q_{a}\right) ^{2}+\left( q^{\prime }-q_{a}\right) ^{2}\right) /2\xi
%_{a}^{2}}$\ with  $\xi _{a}^{2}=\frac{\hbar }{2M_{ph}\omega _{a}} $ whereas
%for $\Delta \mu >>\omega_a$ 
%$r_a (q,q^{\prime })=\frac{1}{\zeta _{a}\sqrt{\pi }} \delta_{qq'}
%e^{-\left( \left(q-q_{a}\right) ^{2} /2\zeta_{a}^{2}\right)}$ 
%with  $\zeta^2=\frac{\Delta \mu} {M_{ph}\omega_a^2}
%\frac{a_{L}a_{R}}{\left( a_{L}+a_{R}\right)
%^{2}}\left( 1+\mathcal{O}g^{2}\right) $.    The peak width vanishes
%if $M_{ph} \rightarrow \infty$ and $\Delta \mu \rightarrow 0$. 

Fixing the $\rho_a$ requires 
consideration of the exponentially small processes neglected above. 
Within the present formalism we find two: 
quantal (finite $M_{ph}$)  effects which lead to tunnelling
through the barrier connecting the minima, and 
diffusion (finite $\Delta \mu$) effects which produce
motion along the pseudoenergy surface. Diffusion 
may be analysed using the analogy between $\Delta \mu$
and temperature mentioned above; it leads \cite{Kramers40} to motion
between minima with rate constant $R_{diff}$ of order 
$\ln R_{diff}\approx \frac{H}{\Delta \mu }$
if $\Delta \mu >>\omega_a$ and to a much smaller rate in the opposite limit.
The standard equilibrium estimate of the rate due to 
tunnelling processes gives
$lnR_{tun}\approx -2\sqrt{\frac{HK_{a}\left( q_{1}-q_{2}\right) ^{2}}{\omega
_{a}}}\sim \frac{H}{\omega _{a}}$; corrections  become
important only for $\Delta \mu \sim \omega_a$. Thus we expect roughly that for $\Delta
\mu >\omega _{a}$, the diffusion process dominates.
\begin{figure}[th]
\includegraphics[width=3.0in]{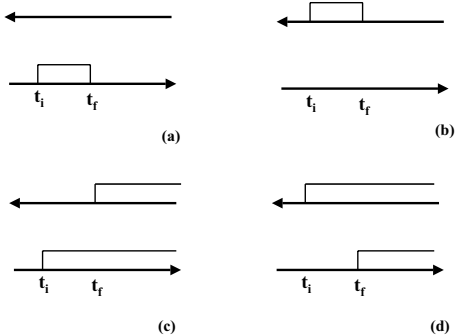}
\caption{Instanton processes entering equation of motion of density matrix. }
\label{fig:instantons}
\end{figure}

We now consider in
more detail the interesting quantum limit $\Delta \mu <\omega _{a}$ for
large barrier height $H$. In this limit one would like to restrict attention to 
paths for which $q(t)$ is almost always near one of the minima, with occasional
tunnelling events in which $q$ shifts from one minimum to another.  The tunnelling
amplitude $R_{tun}$ is exponentially small and (for small enough $\Delta \mu$)
is well approximated by its equilibrium value. However, in the 
real time path integral  the smallness of $R_{tun}$ arises
as the nearly complete cancellation of a sum of many
paths with large but oscillating amplitudes.  We argue that for small $\Delta \mu$
the result of performing this complicated sum (which we do not treat directly)
may be represented by the 'instanton' (kink) vertices shown in Fig \ref{fig:instantons}
with amplitude $R_{tun}$.  A kink or antikink comes with a factor $ i$ and a sign 
$\sigma= \pm 1$ determined
by the contour it is on. We find (see below) that the action of a single kink
or antikink is infinite, so each kink must be followed by an antikink.  The
path integral Eq \ref{rhooft} thus becomes
\begin{eqnarray}
\rho_a(t)=\rho_a(t)&-R_{tun}^2\int_{t_0}^tdt_fdt_i  
({\cal R}(t_f,t_i;\left\{q_{ca} \rightarrow q_{ca}\right\})\rho_a(t_0) \nonumber \\
&-{\cal R}(t_f,t_i;\left\{q_{ca} \rightarrow q_{cb}\right\}\rho_b(t_0))+{\cal O}R^4
\label{rhooftkink}
\end{eqnarray}
where the first term is the sum of the processes shown in panels $a$ and $b$
in Fig. \ref{fig:instantons} the second term is the sum of
the processes shown in $c$ and $d$, and we have denoted explicitly
only the beginning and ending values of the classical component of the field.
%Convergence of the expansion is controlled by the magnitude of $R_{tun}$
%(very small in the large $H$ limit) and the integral over the instanton duration
%$t_{fi}=t_f-t_i$, which is discussed below.

Differentiating Eq \ref{rhooftkink}  with respect to time yields
 \begin{equation}
\frac{d\rho _{a}(t)}{dt}=-\Gamma _{a\rightarrow b}\rho _{a}(t)+\Gamma
_{b\rightarrow a}\rho _{b}(t)  \label{kinetic}
\end{equation}
where to order $R_{tun}^2$ the  scattering rates are 
\begin{equation}
\Gamma_{a\rightarrow a'}=R_{tun}^2Re\int_{0}^{\infty }dt
{\cal R}(t,0;\left\{q_{ca} \rightarrow q_{ca'}\right\})
\label{rates}
\end{equation}
The integral in Eq \ref{rates} yields a quantity $T_0$ with dimension of time;
we may neglect higher order terms  if $R_{tun}T_0<<1$.
%and $%
%\Gamma _{b\rightarrow a}=Re\int_{0}^{\infty }dt\mathcal{R}_{b,a}$
%respectively, with $\mathcal{R}_{ab}(t)$ the time evoluation operator for
%the process shown in panels a or b of Fig. \ref{fig:instantons} with the
%kink from $a$ to $b$ at time $t=0$ and the antikink $b\rightarrow a$ at time 
%$t>0$ and $\mathcal{R}_{b,a}$ the time evolution operator for the processes
%shown in panels $c$ or $d$. The signs of the rates arise mathematically
%because a kink beginning at time $t_{i}$ on one contour and ending at a time 
%$t_{f}$ on a possibly different contour comes with a factor $%
%i^{2}a(t_{i})a(t_{f})$. 
To evaluate Eq \ref{rates} it suffices to 
approximate the potential in Eq \ref{glamdef} by the 'telegraph'
form shown in Fig \ref{fig:instantons}; the  equations may then be solved
by standard methods.  In equilibrium a crucial role is played by the phase shift $%
\delta =\tan ^{-1}\left( \frac{\left( a_{L}+a_{R}\right) V}{1-bV}\right) $ 
\cite{Nozieres69} with $V=\lambda\left( q_{b}-q_{a}\right) $ and $a_{L,R}$ and $b$
given by Eqs \ref{grdef},\ref{gkdef} at $q=q_{a}$. 
As shown by Ng \cite{Ng96}, out of equilibrium,
the behavior is described by complex phase shifts $\delta _{L,R}$ with%
\begin{equation}
\delta _{L,R}=\tan ^{-1}\left( \frac{a_{L,R}V}{1-bV-ia_{R,L}Vsgn\left( \mu
_{L,R}-\mu _{R,L}\right) }\right)   \label{deltal}
\end{equation}
Note $\delta _{L}+\delta _{R}=\delta $.  We find

\begin{equation}
ln{\cal R}=(C_{eq}(\overline{\mu })-i\delta
E_{neq}+\delta C_{dec}+\delta C_{orth})t_{fi}
\label{R}
\end{equation}

Here $\delta E_{neq}=\left( \delta _{L}^{\prime }-\delta _{R}^{\prime }\right)
\Delta \mu $ is the change in electronic energy arising from the imbalance
in chemical potential.  $C_{eq}(\overline{\mu })-i\Delta E_{neq}$ is the equilibrium result 
evaluated at the mean chemical potential plus the nonequilibrium 
energy correction; it oscillates
at frequency $\Delta E$, providing convergence of the integral in 
Eq \ref{rhooftkink} when $\Delta E \neq 0$. 
The imaginary parts $\delta ^{\prime \prime }$ of
the phase shifts give rise to a time decay ('decoherence') which is the
manifestation, in the present formalism, of the decoherence
introduced on semiphenomenological grounds by Rosch et. al. \cite{Rosch03}.
and ensures that  out of equilibrium the integral in Eq \ref{rhooftkink} converges
even when
$\Delta E +\Delta E_{neq}=0$.  We find
\begin{equation}
\delta C_{dec}=-\frac{\left( \delta _{L}^{^{\prime \prime }}-\delta
_{R}^{^{\prime \prime }}\right) }{2\pi }\Delta \mu \phi _{2}(\Delta
\mu t_{fi})  \label{decoh}
\end{equation}%
At short times $\Delta \mu t_{fi}<<1$, $\delta C_{dec}t_{fi}\sim -\left( \Delta
\mu t_{fi}\right) ^{2};$ at long times $\delta C_{dec}t_{fi}\sim -\left\vert
\Delta \mu t_{fi}\right\vert $.  At intermediate times an analytical
solution is not available and $\phi_2$ must be computed
perturbatively or numerically. We find $\phi _{2}(x)=\left( \frac{2%
}{\pi }\right) \left( Si(x)-\frac{1-\cos (x)}{x}\right) $.

$\delta C_{orth}$expresses the change in orthogonality effects arising
because at nonvanishing $\Delta \mu $ there is destructive interference
between the left and right leads \cite{Ng96}:%
\begin{equation}
\delta C_{orth}t_{fi}=\frac{2Re\delta _{L}\delta _{R}}{\pi ^{2}}\psi
_{1}(\Delta \mu t)+\frac{2iIm\delta _{L}\delta _{R}}{\pi ^{2}}\psi
_{2}(\Delta \mu t)  \label{orthog}
\end{equation}
with $\psi
_{1}(x)=\gamma _{e}-Ci(x)+\ln (x)$ and $\psi
_{2}(x)=\frac{2}{\pi}\int_{0}^{1}du\sin (ux)\frac{\left[ \left( 1-u\right) \ln (1-u)+u\ln
u\right] }{u^{2}}$.

\begin{figure}[tbp]
\includegraphics[width=3.0in]{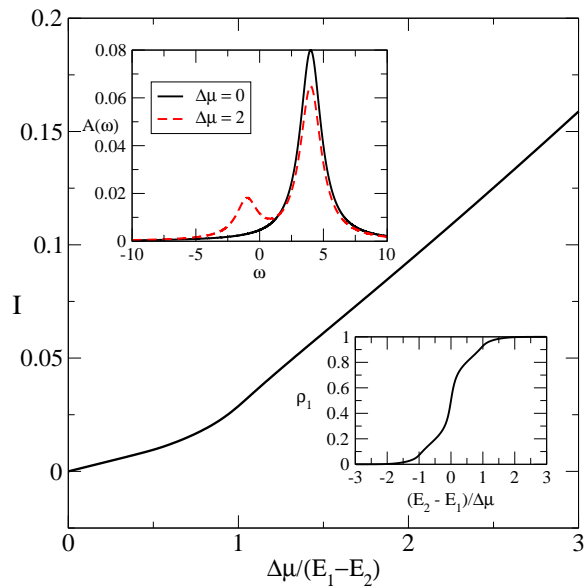}
\caption{\textit{Main panel}: Variation of current with chemical potential
difference $\Delta\protect\mu$ at fixed pseudoenergy difference $%
E_{1}-E_{2}$ calculated for parameters $\lambda q_{1}=4 \Sigma^{\prime\prime}$
(resonance well above fermi level) and $\lambda q_{2}=-\Sigma^{\prime\prime}$
(resonance slightly below fermi level) and equal
coupling to left and right leads. \textit{Upper Inset:} Spectral functions
corresponding to equilibrium state (solid curve--only minimum "1" occupied)
and strongly nonequilibrium state (dashed curve; incoherent superposition of
the $G$'s corresponding to the two minima). {\it Lower Inset}: 
Variation with pseudoenergy difference of density matrix element corresponding
to states near pseudoenergy minimum 1, computed for parameters 
used in main panel. }
\label{fig:GandI}
\end{figure}

We have used the perturbative expressions for the crossover functions to
evaluate $\Gamma$ (Eq \ref{rates}).
%The scattering out and scattering in rates for level $a$ are $\Gamma
%_{a\rightarrow b}=Re\int_{0}^{\infty }dt\mathcal{R}_{ab}(t)$ and $%
%\Gamma _{b\rightarrow a}=Re\int_{0}^{\infty }dt\mathcal{R}_{b,a}$
%respectively, with $\mathcal{R}_{ab}(t)$ the time evoluation operator for
%the process shown in panels a or b of Fig. \ref{fig:instantons} with the
%kink from $a$ to $b$ at time $t=0$ and the antikink $b\rightarrow a$ at time 
%$t>0$ and $\mathcal{R}_{b,a}$ the time evolution operator for the processes
%shown in panels $c$ or $d$. The signs of the rates arise mathematically
%because a kink beginning at time $t_{i}$ on one contour and ending at a time 
%$t_{f}$ on a possibly different contour comes with a factor $%
%i^{2}a(t_{i})a(t_{f})$. 
In equilibrium and at $\beta_j =\infty $ we find that
only scattering from the higher energy to the lower energy extremum occurs
(the 'up-scattering rate' vanishes); under nonequilibrium conditions an
upscattering rate appears: of order 
$\left(\frac{\Delta \mu}{\Delta E}\right)^{(\frac{\Delta E}{\Delta \mu})}$ if 
$\frac{\left\vert \Delta E \right\vert }{\Delta \mu }>>1$ but  of the order of the
down-scattering rate as $\frac{\left\vert \Delta E \right\vert }{\Delta
\mu }\rightarrow 0$.  The change in the  weight $\rho_1$ for one of the two minima
as $\Delta \mu /\Delta E$ is varied 
is shown in the lower inset of Fig \ref{fig:GandI}. A similar result was found by
Parcollet and Hooley in a 'pseudofermion' diagrammatic study of the
nonequilibrium Kondo problem \cite{Parcollet02}.

The semiclassical Greens function follows as:
\begin{equation}
{G}_{R}(\omega )=\sum_{a}\frac{\rho _{a}\left( (\Delta \mu \right) }{%
\omega -\epsilon _{0}-gq_{a}-\Sigma _{R}(\omega )}  \label{GRinc}
\end{equation}%
The main panel of Fig \ref{fig:GandI} shows the current computed in the standard
way by inserting
the Green function given in Eq. \ref{GRinc} into Eq 28 of Ref \cite{Mitra04}.
The inset shows the evolution
of the spectral density $A(\omega)=ImG_R(\omega)$. 

The minimum condition $\partial \Phi /\partial q=0$ is equivalent to the
Hartree-Fock equations discussed in recent literature \cite%
{Alexandrov02,Gogolin04}, but in these works it is assumed that at each set
of parameter values, only one of the minima is occupied ($\rho _{a}=0$ or $1$%
);  preparation 
conditions are argued \cite{Alexandrov02,Gogolin04} to determine 
the state of the system, leading to multistability and switching,
in contradiction to the results shown in Fig \ref{fig:GandI}.
Although the bistability
discussed by \cite{Alexandrov02,Gogolin04} does not occur, 
the slow dynamics governing equilibration
between the minima will lead to  'telegraph' noise in the current.

Hamann \cite{Hamann70} shows that interacting electron
problems such as the Anderson model lead, after Hubbard-Stratonovich
transformation, to a model very like that studied here, but with the role of
the phonon field played by the spin part of the decoupling field. Our
analysis  carries over directly to these problems, providing 
a different derivation \cite{Mitra04b} of the generally
accepted results \cite{Parcollet02,Rosch03} that nonequilibrium effects
suppress the formation of the Kondo resonance, and that the 
nonequilibrium Kondo effect  is fundamentally a weak coupling problem.

\textit{Acknowledgements:} We thank the Columbia University Nanocenter and
NSF-DMR0338376 for support, and L.Glazman for very helpful conversations.

\end{document}